\documentclass[twocolumn]{aastex62}
\usepackage{graphicx}
\usepackage{color}
\usepackage{amsmath}
\usepackage{comment}
\usepackage{lineno}
\usepackage[first=0,last=9]{lcg}

\usepackage{array,multirow}

\definecolor{LightCyan}{rgb}{0.88,1,1}

\usepackage{appendix}
\usepackage{tabularx}
\usepackage{longtable}
\usepackage{enumerate}
\usepackage{enumitem}

\definecolor{pink}{RGB}{255, 20, 147}

\definecolor{carminered}{rgb}{1.0, 0.0, 0.22}

\definecolor{amber}{rgb}{0.95, 0.8, 0.2}

\definecolor{blue}{rgb}{0.01, 0.01, 0.98}

\definecolor{byzantine}{rgb}{0.74, 0.2, 0.64}

\definecolor{amethyst}{rgb}{0.6, 0.4, 0.8}

\definecolor{blue-violet}{rgb}{0.54, 0.17, 0.89}

\definecolor{comment}{RGB}{166, 38, 164}

\begin{document}

\title{\bf Radio Constraints on $r$-process Nucleosynthesis by Collapsars}

\author{K.H. Lee}
\affiliation{Department of Physics, University of Florida, PO Box 118440, Gainesville, FL 32611-8440, USA}
\author{I. Bartos}
\thanks{imrebartos@ufl.edu}
\affiliation{Department of Physics, University of Florida, PO Box 118440, Gainesville, FL 32611-8440, USA}
\author{A. Cook}
\affiliation{Department of Physics and Astronomy, Texas Tech University, Box 1051, Lubbock, TX 79409-1051, USA}
\author{A. Corsi}
\affiliation{Department of Physics and Astronomy, Texas Tech University, Box 1051, Lubbock, TX 79409-1051, USA}
\author{Z. M\'arka}
\affiliation{Department of Physics, Columbia University in the City of New York, New York, NY 10027, USA}
\author{G.C. Privon}
\affiliation{National Radio Astronomy Observatory, 520 Edgemont Rd, Charlottesville, VA 22903, USA}
\author{S. M\'arka}
\affiliation{Columbia Astrophysics Laboratory, Columbia University in the City of New York, New York, NY 10027, USA}

\begin{abstract}
The heaviest elements in the Universe are synthesized through rapid neutron capture ($r$-process) in extremely neutron rich outflows. Neutron star mergers were established as an important $r$-process source through the multi-messenger observation of GW170817. Collapsars were also proposed as a potentially major source of heavy elements; however, this is difficult to probe through optical observations due to contamination by other emission mechanisms. Here we present observational constraints on $r$-process nucleosynthesis by collapsars based on radio follow-up observations of nearby long gamma-ray bursts. We make the hypothesis that late-time radio emission arises from the collapsar wind ejecta responsible for forging $r$-process elements, and consider the constraints that can be set on this scenario using radio observations of a sample of Swift/BAT GRBs located within 2\,Gpc. No radio counterpart was identified in excess of the radio afterglow of the GRBs in our sample, limiting the collapsar $r$-process contribution to $\lesssim0.2$\,M$_\odot$ under the models we considered, with constant circum-merger densities giving more stringent constraints. While our results are in tension with collapsars being the majority $r$-process production sites, the ejecta mass and velocity profile of collapsar winds is not yet well modeled. As such, our results are currently subject to large uncertainties, but further theoretical work could greatly improve them.

\end{abstract}

\keywords{gamma ray burst, collapsar, $r$-process nucleosynthesis, radio observations}

\section{Introduction}

The remarkable multi-messenger discovery of the neutron star merger GW170817 by the LIGO \citep{2015CQGra..32g4001L} and Virgo \citep{2015CQGra..32b4001A} gravitational-wave observatories and their partners has marked the start of a new era in astrophysics \citep{PhysRevLett.119.161101}. Among others, it confirmed that neutron star mergers eject mildly relativistic, neutron rich matter. This ejecta can produce the heaviest elements known in nature through rapid neutron capture ($r$-process) nucleosynthesis \citep{2017Sci...358.1570D,2017ApJ...848L..19C,2017ApJ...850L..37P,2017Natur.551...67P}. 

Based on the ejecta mass from GW170817, neutron star mergers could be the main source of heavy $r$-process elements in the Universe. Nevertheless, theoretical and observational uncertainties remain \citep{2020ApJ...900..179K,2019EPJA...55..203S,2019Natur.569...85B}. 

Collapsars---rapidly rotating massive stars whose iron-core collapse produces a black hole \citep{1999A&AS..138..499W}---have also been proposed as a major source of heavy $r$-process elements \citep{2013IJMPE..2230022N,2015A&A...582A..34N,2019Natur.569..241S,2020arXiv201015837B}. This possibility could help explain the observed early enrichment of dwarf galaxies \citep{2016Natur.531..610J} and the long-term chemical enrichment of the Milky Way \citep{2017ApJ...836..230C,2018IJMPD..2742005H}, albeit their low rate is in tension with meteoritic $r$-process abundances \citep{2019ApJ...881L...4B}. Fallback accretion onto the newly formed black hole in collapsars could drive winds similar to those observed in neutron star mergers, except with a larger ejecta mass due to more available mass in collapsar accretion disks. This wind outflow, hereafter ''collapsar wind ejecta", could become highly neutron rich due to electron capture reactions on protons \citep{2019Natur.569..241S}, enabling it to robustly synthesize even heavy $r$-process elements.

So far, there is no direct evidence of $r$-process nucleosynthesis by collapsars. Such an observation is difficult since most collapsars are much farther than GW170817, and because optical emission from decaying $r$-process elements, which was observed for GW170817, can be outshone by supernovae ignited by the collapsars \citep{2020MNRAS.499.4097Z}. Hence it has been difficult to confirm, or rule out, collapsars as major $r$-process production sites.

Collapsar wind ejecta could produce late-time, slowly-evolving radio flares driven by the outflow's interaction with the surrounding medium. This radio emission is analogous to the late-time, slowly-rising radio flares expected in the case of neutron star mergers \citep{2011Natur.478...82N,2020arXiv200805330G,2019MNRAS.485.4150B}, which may have already been detected \citep{2020ApJ...902L..23L,2021arXiv210402070H}. Collapsars, however, may be more favorable for the detection of radio flares: (i) since they are rarer than neutron star mergers, they individually need to eject more matter than a neutron star mergers to explain the observed abundance of $r$-process elements in the Universe, and (ii) they are typically found in a denser interstellar medium (ISM) than neutron star mergers, potentially leading to brighter radio flares.

The emergence, or lack, of a late-time, slowly-evolving radio flare could provide observational evidence for or against a mildly-relativistic collapsar wind ejecta and could help constrain $r$-process nucleosynthesis by collapsars. This picture may nevertheless be complicated by other outflows from collapsing massive stars, including the beamed, relativistic ejecta that produces gamma-ray bursts (GRBs), and the supernova explosion that ejects a part of the stellar envelope. The emergence of GRBs poses a limited challenge to the interpretation of collapsar radio flares, as radio emission from relativistic outflows that produce the GRBs peak much earlier. In addition, the properties of the relativistic ejecta can be inferred from the observation of the GRB and its afterglow, therefore it can be accounted for in the interpretation of an observed long-term radio signal. 

The effect of supernovae on the observability of collapsar radio flares is less clear. Supernovae contain much more kinetic energy than GRB jets, albeit they expand at a much lower velocity ($\lesssim0.1$c). This low velocity means that the supernova ejecta leads to a radio flare that peaks only decades after the onset of the supernova \citep{2015MNRAS.454.1711B}. While long-lasting radio emission peaking at late times has been observed in broad-lined type Ic supernovae (the same type of supernovae associated with GRBs) strongly interacting with the circumstellar medium (e.g., PTF11qcj; \citealt{2016ApJ...830...42C}), no radio rebrightening associated with GRB-supernova events has been identified so far \citep{2019ApJ...872...28P}. However, exceptions are possible. Thus, in case of a detection compatible with a collapsar radio flare, expectations from an associated supernova needs to be examined. It will also be important to seek further broad-band radio follow-up to characterize the spectral properties of the radio emission.

Some observed long GRBs are detected without a coincident supernova even with very deep observations \citep{2020arXiv201006612E}. There are also reports of the disappearance of a red supergiant after a recent outburst, consistent with a failed supernova \citep{2020arXiv200715658B}. Therefore, it is possible that some collapsars do not produce supernovae. If a GRB is observed from the collapsar, it is also possible that the line of site from the collapsar is ``cleared" from the exploding stellar envelope, and therefore the collapsar wind ejecta also leaves without significant interference.

We investigated the observational constraints on the collapsar wind ejecta that could be connected to the synthesis of $r$-process elements. We considered detected nearby long GRBs and placed limits on their ejecta properties using existing radio follow-up observations. 

\section{Gamma-ray burst sample}
\label{sec:GRB}

To construct a sample of promising targets, we identified long GRBs that were detected by the Burst Alert Telescope (BAT) on-board the Neil Gehrels \textit{Swift} Observatory \citep{2013ApJS..209...14K}. Swift-BAT provides accurate localization for follow-up observations. We only considered publicly available afterglows and reconstructed distances. We down-selected this sample to GRBs within $2$\,Gpc, which could be the brightest sources due to their vicinity. We found 13 GRBs that satisfied these conditions. As we were interested in possible follow-up observations with the Karl G. Jansky Very Large Array (VLA), we discarded 2 GRBs that fell outside the sky region accessible to the VLA (GRBs 180728A and 060614). The remaining 11 GRBs and their properties are listed in Table \ref{table:Host galaxies}. 

\begin{table*}
	\centering
	\begin{tabular}{lccccccccccc}
		\hline
GRB    & R.A.    & Dec.   & err. & SN & $n$ & dist. & Limit & Freq & obs. date & GCN & Ref. \\ 
name   & hh:mm:ss & dd:mm:ss & $''$ &  & cm$^{-3}$ & Mpc & mJy & GHz & & & \\\hline        \hline

191019A & 22:40:5.93 & -17:19:40.9 & 2.4 & - & 0.01 & 1293 & - & - & - & 26043 & - \\\hline
190829A & 02:58:10.57 & -08:57:28.6 & 2.0 & - & 0.0005 & 373 & - & - & - & 25552 & - \\\hline 
171205A & 11:09:39.37 & -12:35:20.1 & 2.3 & 2017iuk & n/a & 168 & 0.5 & 3 & 2020.10.10 & 22179 & - \\\hline 
161219B & 06:06:51.36 & -26:47:29.9 & 1.7 & 2016jca & 2.26 & 726 & 0.6 & 3 & 2019.07.04 & 20297 & - \\\hline
150727A & 13:35:52.42 & -18:19:32.7 & 1.8 & - & 1.86 & 1688 & 0.5 & 3 & 2019.06.30 & 18079 & - \\\hline 
111225A & 00:52:37.34 & +51:34:17.6 & 2.2 & - & 0.01 & 1589 & 0.02 & 6 & 2017.09.15 & 12724 & - \\\hline 
061021 & 09:40:35.87 & -21:57:07.2 & 5 & - & $10^{-3}$ & 1898 & 0.5 & 3 & 2019.06.29 & 5746 & [1]\\\hline
060505 & 22:07:4.5 & -27:49:57.8 & 4.7 & - & 1 & 422 & 0.5 & 3 & 2020.10.25 & 5081 & [2]\\ \hline
060218 & 03:21:39.7 & +16:52:01.8 & 3.6 & 2006aj & 0.002 & 150 & 0.5 & 3 & 2019.03.24 & 4786 & [1]\\\hline
051109B & 23:01:50.35  & +38:40:49.6 & 4 & - & 0.01 & 377 & 0.4 & 3 & 2019.04.13 & 4226 & [1] \\ \hline
050826 & 05:51:01.3 & -02:38:41.9 & 6 & - & 600 & 1589 & 0.5 & 3 & 2020.08.15 & 3889 & [1] \\ \hline
	\end{tabular}
	\caption{\label{table:Host galaxies} {\bf List of Swift-BAT long GRBs with distance $<2$\,Gpc.)}. For GRBs for which a best estimate of $n$ was not found in the literature, we adopted the fiducial value $n=0.01$\,cm$^{-3}$. Dist.: luminosity distance is converted by using Planck 2018 results \citep{collaboration2018planck}. GCN: number for GRB localization. Ref.: where $n$ and distance are taken from: [1] \cite{ryan2015gamma}, [2] \cite{xu2008search}.}
\end{table*}

\section{Ejecta and Radio Emission Model} \label{sec:emissionmodel}

We modeled radio emission due to the interaction of the collapsar wind ejecta and the surrounding medium following the prescription of \cite{2013MNRAS.430.2121P}. The velocity profile of collapsar wind ejecta currently uncertain (e.g., \citealt{2020MNRAS.499.4097Z}). Here we considered two possible velocity profiles. As our fiducial model, we adopted the velocity distribution obtained by \cite{fernandez2019long} who carried out a three dimensional, general-relativistic magnetohydrodynamic (GRMHD) simulation of black hole accretion disks. The simulation was continued long enough to achieve the completion of mass ejection from the disk. This study was motivated by neutron star mergers, however the black hole--accretion disk configuration is similar in the collapsar case. For comparison we also considered a second model in which the velocity profile of the disk wind follows a power law distribution. For this distribution, we adopted a power-law index of $\alpha = 4.5$ based on the results of \cite{2021arXiv210402070H}, and limited the velocity to above $v_{\rm min} = 0.35c$, similar to that of \cite{2021arXiv210402070H}. 

Our standard assumption was that the interstellar medium near the collapsar is uniform, with baryon number density $n$, unless the available afterglow fit for the GRB indicated a wind-like density profile. For uniform density, when the ejecta expands out to a radius $R$, it will accumulate $4/3\pi R^3n+M_{\rm ej}(\beta)$ mass from the surrounding medium. Here, $M_{\rm ej}(\beta)$ is the part of the ejecta mass with initial velocity $\geq\beta$. This can be used to compute $\beta=\beta(R)$ using energy conservation \citep{2013MNRAS.430.2121P}:
\begin{equation}
M(R)(\beta c)^2 \approx E(\geq\beta).
\label{eq:E}
\end{equation}
where $E(\beta)$ is the total kinetic energy of the part of the ejecta mass with initial velocity $\geq\beta$. We computed $\beta(R)$ by solving Eq. \ref{eq:E} numerically, which was then used to obtain $R(t)=\int_{0}^{t}\beta(R)^{-1}dR$ where $t$ is time since the start of the outflow.

The typical electron synchrotron frequency in the shock produced by the ejecta medium interaction is \citep{2013MNRAS.430.2121P}
\begin{equation}
\nu_{\rm m}(t) \approx \textup{1\,GHz}\cdot n_{0}^{\frac{1}{2}}\epsilon_{\rm B, -1}^{\frac{1}{2}}\epsilon_{\rm e, -1}^{2}\beta(t)^{5}
\end{equation}
while the self-absorption frequency is
\begin{equation}
\nu_{\rm a}(t) \approx \textup{1\,GHz}\cdot R_{17}(t)^{\frac{2}{p+4}} n_{0}^{\frac{6+p}{2(p+4)}}\epsilon_{\rm B, -1}^{\frac{2+p}{2(p+4)}}\epsilon_{\rm e, -1}^{\frac{2(p-1)}{p+4}}\beta(t) ^{\frac{5p-2}{p+4}}.
\end{equation}
Here, $p$ is the electron Lorentz factor power law index. The parameters $\epsilon _{\rm B}$ and $\epsilon _{\rm e}$ are the fractions of the total internal energy of the shocked gas carried by the magnetic fields and electrons. For GRBs where an afterglow fit was not available, we used {\it Afterglowpy} package \citep{ryan2020gamma} to generate afterglow light curves for fitting. We used the trimmed data set of UKSSDC XRT light-curve which is available from the Swift-XRT catalog. Each time of generation of lightcurve, Flux density obtained from {\it Afterglowpy} were integrated with spectral function over XRT frequency range (0.3 keV - 10 keV) to obtain the Flux. Then, we performed the MCMC analysis by using the parallel tempering method where the ensembles are run through the likelihoods of $p(D/\Theta)^{1/T}$ where T is the temperature.
$\Theta = (\theta_{v}, E_{iso}, \theta_{c}, n, p, \varepsilon _{e}, \varepsilon _{B})$
Using {\it Afterglowpy}, we adopted top-hat jet-like afterglow structure as the example. For data fitting, we used {\it ptemcee} package \citep{vousden2016dynamic,foreman2013emcee} to perform parallel tempering ensemble MCMC method, and the bound on uniform prior for $\log_{10}n$ and p is set to be [-5,3] and [2,3], respectively. Initial reference for walkers was set to be random values within [0, 0.02]. From this reference point, we sampled 200 walkers for 20 temperatures each. We first generated 1000 iterations of 'burn-in' process, then we sampled the afterglow lightcurves with 1000 iterations.
If the data points were not sufficient to adequately constrain all parameters, we adopted the fiducial values $p=2.2$ and $n=0.01$ and $\epsilon _{\rm B}=\epsilon _{\rm e}=0.1$.

For the relevant observing frequency $\nu_{\rm obs}$ and time in our analysis, we have $\nu_{\rm obs}>\nu_{\rm a}>\nu_{\rm m}$, for which the expected flux can be written as
\begin{equation}
F_{\rm obs}(t) = 500\,\mu\mbox{Jy}\,R(t)_{17}^3 n_{-1}^{3/2} \beta(t)^{1} d_{27}^{-2} \left(\frac{\nu_{obs}}{\nu_{m}(t)}\right)^{{-\frac{p-1}{2}}}.
\end{equation}

\section{Radio constraints from observations}
\label{sec:constraints}

We examined observations in three radio surveys: the Faint Images of the Radio Sky at Twenty Centimeters (FIRST) survey conducted from 2009 to 2011 over the south Galactic hemisphere with a typical rms sensitivity of 0.15\,mJy \citep{1995ApJ...450..559B}, and two observing epochs of the Karl G. Jansky Very Large Array Sky Survey (VLASS) with typical rms sensitivity of 120\,$\mu$Jy \citep{2020PASP..132c5001L}. We utilized the {\it Photutils} package \citep{larry_bradley_2020_4044744} to estimate the background noise of each GRBs' VLASS Quick-view image above 3$\sigma$ after masking possible radio sources. 

In addition to the above surveys, GRB 111225A was also observed with a dedicated VLA follow-up in 2017 \citep{eftekhari2020late}. 

Only one of the considered GRBs had a reported radio counterpart: GRB 171205A. \cite{leung2021search}  reported a radio afterglow candidate associated
with GRB 171205A, detectable at least until 900 days post-burst. Comparing to early afterglow detections they found that all radio observations can be explained by afterglow emission propagating in a wind medium and with an unusually high electron Lorentz factor power law index $p\approx2.8$. 

For each GRB in our sample, the time of and obtained radio flux limit from the most constraining observation is presented in Table. \ref{table:Host galaxies}. In the case of GRB 171205A, the Table presents the latest non-detection obtained after the reported detection in \cite{leung2021search}.

\section{Constraints on the ejecta mass}

\begin{figure}
\centering
\includegraphics[angle = 0, scale = 0.47]{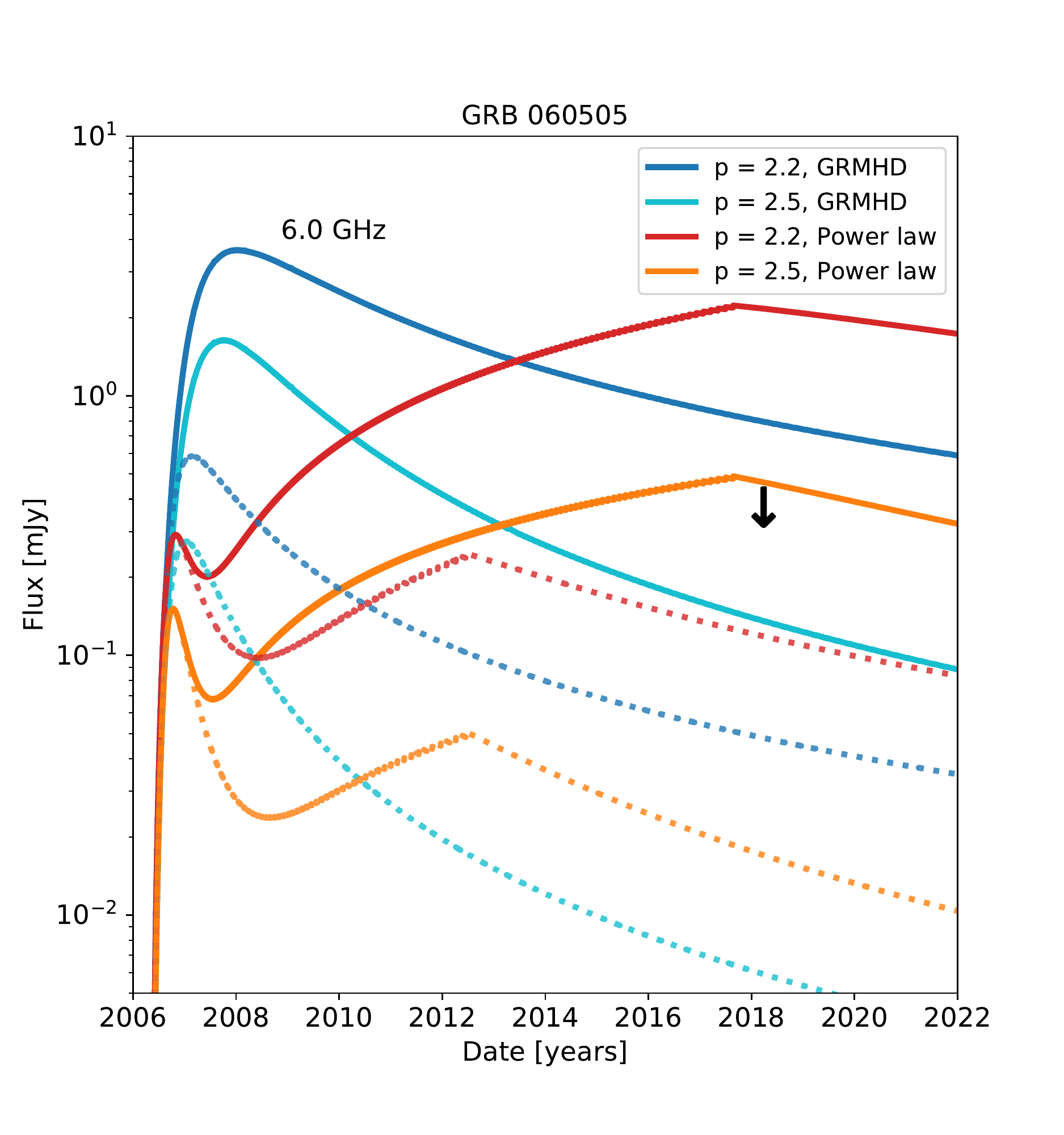}
      \caption{{\bf Simulated radio light curve from the collapsar wind ejecta of GRB 060505 with example parameters.} We considered two possible ejecta velocity profiles, GRMHD and "Power law" (see legend and text), two possible values for the electron Lorentz factor power law index $p$ (see legend), and two possible ejecta masses: $10^{-1}$\,M$_\odot$ (solid) and $10^{-2}$\,M$_\odot$ (dotted). Light curves are at $6\,$GHz. Also shown is an observational constraint obtained by VLA in 2018 (black arrow). The early first peak in the "Power law" case is due to the GRB afterglow, while in the "GRMHD" case the collapsar wind ejecta is much brighter, making the afterglow's effect negligible.
    \label{fig:GRB060505}}
\end{figure}

We used the obtained radio constraints to limit the mass of the collapsar wind ejecta from each GRB. We computed the expected radio flux at the time of the observation for the GRBs for ejecta masses in the range $10^{-3}-2$\,M$_\odot$. 

Out of the GRBs considered, we found that one, GRB 060505, is inconsistent with ejecta mass $> 0.15$\,M$_\odot$ ( $> 0.05$\,M$_\odot$) for our fiducial model with $p=2.5$ ($p=2.2$), while we found that the other GRBs do not constrain the ejecta mass. To demonstrate this constraint and the expected uncertainties due to our model, we show possible radio light curves at 6\,GHz in Fig. \ref{fig:GRB060505}. We see that the velocity profile and different choices for $p$ can significantly alter the light curve.

For GRB 171205A, instead of our fiducial model of constant circumburst density, we considered a wind model with density decreasing with $r^{-2}$. For this case we adopted the density obtained by \cite{leung2021search} through their fit of the afterglow emission of GRB 171205A. With this model, assuming a terminal shock radius of 1\,pc (0.01\,pc) beyond which the medium density is taken to be $1$\,cm$^{-3}$ , we could constrain the ejecta mass to $\lesssim0.5$\,M$_\odot$ ($\lesssim0.05$\,M$_\odot$). 

The flux detected for GRB 171205A \cite{leung2021search} prior to the latest non-detection reported here does not meaningfully constrain the possible collapsar ejecta mass. If we assume that the latest detection at $\sim900$ days after the GRB is fully from the collapsar ejecta, i.e. not from the afterglow, we find that the ejecta mass needs to be $>1$\,M$_\odot$, i.e. a mass less than this cannot be ruled out by the observation.

Long GRBs occur about 4 times less frequently as short GRBs. This means that collapsars need to have a wind ejecta mass of more than about $0.1-0.2$\,M$_\odot$ to be the major $r$-process sources in the Universe, if we adopt the ejecta inferred from GW170817 as typical for neutron star mergers. Even higher ejecta masses (up to $\sim2$\,M$_\odot$) are possible under the assumption that black hole accretion in collapsars and neutron star mergers is directly proportional to the radiated $\gamma$-ray energy  \citep{2019Natur.569..241S}.  Our fiducial model with constant density circum-merger medium constrains the ejecta mass is at this threshold. We conclude that with future observations, and the reduction of model uncertainties, radio observations may be a powerful direct means to establish the $r$-process contribution of collapsars.

\section{Conclusions} \label{sec:methods}

We searched for the radio signature of collapsar wind ejecta for 11 nearby ($<2$\,Gpc) well localized GRBs ($<5''$) consider. We analyzed radio limits from the FIRST and VLASS surveys and a 2017 follow-up observation of GRB 111225A. We found no coincident radio signal beyond the afterglow of the GRBs, which we used to constrain the mass of collapsar wind ejecta. 

Assuming a constant-density interstellar medium around the collapsars, we found that ejecta with $\gtrsim0.2$\,M$_\odot$ mass are inconsistent with observations for our fiducial model. Considering GRB 171205A and a wind profile for the circum-burst density \citep{leung2021search}, we obtain a mass constraint of $\lesssim0.5$\,M$_\odot$ ($\lesssim0.05$\,M$_\odot$) for 1\,pc (0.01\,pc) assumed shock termination radius.  

Uncertainties in our result include limited information on the circum-burst medium density, uncertainty in some of our model parameters ($p$, $\epsilon_{\rm e}$, $\epsilon_{\rm B}$), and possible interference by the stellar matter ejected by the supernova. Further, the ejecta mass and velocity profile of collapsar winds is not yet well modeled. Nonetheless, our results show that long-term radio emission can be a viable option to probe $r$-process nucleosynthesis by collapsars that is difficult through other means.

\acknowledgements
We would like to thank Brian Metzger for the useful feedback. The National Radio Astronomy Observatory is a facility of the National Science Foundation operated under cooperative agreement by Associated Universities, Inc. A.C. acknowledge support from NSF CAREER award \#1455090 and NSF award \#1841358. Z.M. and S.M. thank Columbia University in the City of New York its unwavering support. Z.M., S.M. and the Columbia Experimental Gravity group is grateful for the generous support of the National Science Foundation under grant PHY-1708028. I.B. acknowledges the support of the National Science Foundation under grant PHY-1911796 and PHY-2110060, the Alfred P. Sloan Foundation, and the University of Florida.

\bibliography{Refs}
\end{document}